\documentclass[12pt]{iopart}
\usepackage{iopams}
\usepackage{url}
\usepackage{graphicx,color}
\usepackage{verbatim}
\usepackage{bbm}
\usepackage{epstopdf}

\usepackage{hyperref}

\usepackage[T1]{fontenc}
\usepackage{ae,aecompl}

\usepackage{amssymb,amscd}
\usepackage{url}
\usepackage{graphicx}

\newcommand{\cA}{\mathcal{A}}

\newcommand{\cL}{\mathcal{L}}

\newcommand{\cO}{\mathcal{O}}

\newcommand{\cS}{\mathcal{S}}

\newcommand{\EE}{\mathbb{E}}

\newcommand{\HH}{\mathbb{H}}

\newcommand{\PP}{\mathbb{P}}

\begin{document}

\title{Accessibility percolation in random fitness
  landscapes}

\author{Joachim Krug}

\begin{abstract}
The fitness landscape encodes the mapping of genotypes to fitness and provides a succinct representation of possible trajectories followed by an evolving
population. Evolutionary accessibility is quantified by the existence of
fitness-monotonic paths connecting far away genotypes. Studies of 
accessibility percolation use probabilistic fitness landscape 
models to explore the emergence 
of such paths as a function of the initial fitness, the parameters of the 
landscape or the structure of the genotype graph. This chapter reviews 
these studies and discusses their implications for the predictability
of evolutionary processes.
\end{abstract}

\noindent
\small{
\textit{Determining the paths that evolution does not take is as
  important in evolutionary outcomes as shaping those it may pass
  through.} \hfill T.N.~Starr~and~J.W.~Thornton~\cite{JK-Starr2016}

\normalsize

\section{Introduction}

An evolving population traces out a path in the space of genetic
sequences or genotypes. Depending on the level of resolution,
the genotype may be described in terms of nucleotide bases, amino acid
residues or the alleles of genes. It has been recognised for a long
time that sequence spaces are vast, and that only a tiny fraction of
all sequences code for viable phenotypes. This raises the question
of how evolution nevertheless manages to navigate these spaces across
macro-evolutionary distances \cite{JK-Arnold2011,JK-Louis2016}.

An early mathematical formulation of this problem was presented by
John Maynard Smith, who estimated the fraction of functional
proteins from a percolation argument
\cite{JK-MaynardSmith1970}. Assuming that evolution is restricted to
proceed by single amino acid substitutions, he conceptualised the
space of all protein sequences as a network where each protein is
connected to its $n$ one-step mutant neighbours, and a fraction $p$ of
proteins is functional. The expected number of functional
  neighbours of a given sequence is therefore $pn$. It is then plausible (and can be proved
\cite{JK-Gavrilets1997,JK-Reidys1997,JK-Reidys2009}) that a large connected component of functional proteins
exists with high probability if $pn > 1$. Since typically $n \sim
10^3$, only a small fraction of proteins has to be functional to ensure
evolvability over large genetic distances. 

In this chapter we introduce and review a different kind of
percolation problem motivated by evolutionary adaptation. In its most
general form the problem of \textit{accessibility percolation} can be formulated as follows
\cite{JK-Nowak2013}. Consider a graph $G = (V,E)$ where each vertex $x
\in V$ is labelled by a real-valued random number $f_x$ drawn from a
joint continuous
distribution. We call a path between two vertices $x,y$
\textit{accessible} if the random numbers along the path are
monotonically increasing\footnote{The requirement of strict
  monotonicity will be relaxed in Sect.~\ref{JK-Sec:Downhill}.}, and we ask for the probability of existence
of such a path when the distance $d(x,y)$ between the two points (and
the size of the graph) becomes large. The use of the term
  \textit{percolation} in this context is motivated by the fact that
  one studies a certain connectivity property of a random
  structure. A more specific relation to conventional forms of
  percolation will be described below in Sect.~\ref{JK-Sec:Directed}.

The notion of evolutionary accessibility was first
introduced by Daniel Weinreich and collaborators
\cite{JK-Weinreich2005}. In a seminal empirical study they constructed
and characterised all $2^5 = 32$ combinations of 5 point mutations in
the bacterial antibiotic resistance gene TEM-1
$\beta$-lactamase \cite{JK-Weinreich2006}. TEM-1 confers resistance to
ampicillin but has very low activity against the novel antibiotic
cefotaxime. In combination, the 5 point mutations in TEM-1 increase the
baseline resistance against cefotaxime (measured in terms of the concentration up to
which bacterial growth is possible) by a factor of about $10^5$. The
aim of the study was to reconstruct the mutational pathways along
which the highly resistant mutant could arise, assuming that mutations
occur one at a time and that every step has to provide a benefit in
terms of increased resistance\footnote{The conditions for
  these assumptions to hold are summarised in Sect.~\ref{JK-Sec:Dynamics}.}. Since each path corresponds to a
particular order of occurrence of the mutational steps, there are 5! =
120 distinct paths, of which only 18 were found to be
accessible. This observation lead the authors to conclude that
adaptive evolution is more constrained, and hence more predictable,
than previously appreciated.

Along with other related empirical studies
\cite{JK-Lozovsky2009,JK-Poelwijk2007}, the work of Weinreich et al. 
motivated the 
first theoretical investigations of evolutionary accessibility in random fitness
landscapes \cite{JK-Carneiro2010,JK-Franke2011}. Here the term
\textit{fitness landscape} refers to the assignment of
fitness values to genotypes that are connected by mutations 
\cite{JK-deVisser2014,JK-deVisser2018,JK-Fragata2019,JK-Stadler2002,JK-Szendro2013}. 
In terms of the 
general definition of accessibility percolation given above, the graph
of interest in this case is the space of genetic sequences
endowed with the standard Hamming metric, and the random labels
encode fitness or some proxy thereof, such as
antibiotic resistance.
The precise mathematical setting will be
introduced in the next section, and the basic phenomenology will be
explained with particular emphasis on the occurrence of abrupt,
percolation-like transitions in the accessibility properties. Section
\ref{JK-Sec:Trees} reviews accessibility percolation on trees, which sheds
additional light on the role of graph geometry, and some
generalisations of the standard models will be discussed in Sections
\ref{JK-Sec:Correlations} and \ref{JK-Sec:Downhill}. Concluding remarks addressing the relation 
between accessibility and predictability as well as the role of accessible paths in the evolutionary dynamics 
are presented in Sect.~\ref{JK-Sec:Summary}. In
addition to a survey of the literature, some unpublished new results
are reported in Sections \ref{JK-Sec:Multiple} and \ref{JK-Sec:Downhill}. 
In its focus on the structure \textit{of} fitness landscapes, the chapter is
complementary to the contributions of Anton Bovier
\cite{JK-Bovier2019} and Wolfgang
K\"onig \cite{JK-Koenig2019} to this volume, where different types of evolutionary dynamics \textit{on}
fitness landscapes are discussed. 

\section{Accessibility percolation in sequence space}
\label{JK-Sec:SequenceSpace}

\setcounter{footnote}{0}

Genotypes are encoded by sequences $x$ of length $L$ with entries drawn
from an alphabet $\cA = \{0,1,2,..,a-1\}$ of size $\vert \cA \vert =
a$. The elements of $\cA$ will be referred to as \textit{alleles}. 
The Hamming distance between two genotypes $x,y$ is defined by
\begin{equation}
  \label{JK-Hamming}
  d(x,y) = \sum_{i=1}^L (1 - \delta_{\tau_i(x),\tau_i(y)}),
\end{equation}
where $\tau_i(x) \in \cA$ is the $i$'th entry of $x$, that is, the
allele at the $i$'th genetic locus. 
Loci are elements of the locus set $\cL = \{1,2,\cdots,L\}$ with $\vert \cL \vert = L$. 
The sequence space $\cA^L$
endowed with the Hamming metric $d(\cdot,\cdot)$ is the Hamming graph
$\HH_{a}^L$ \cite{JK-Stadler2002}. The binary or biallelic Hamming graphs
$\HH_2^L$ are $L$-dimensional hypercubes, see
Fig.~\ref{JK-Fig1} for illustration.
Fitness values $f_x$ are drawn from a continuous distribution and
assigned independently\footnote{The assumption of independence
  will be dropped in Section \ref{JK-Sec:Correlations}.} to
genotypes \cite{JK-Kauffman1987}. Since accessibility depends only on the rank ordering of
genotypes, the distribution of fitness values does not need to be
specified. In the following we assume without loss of generality that
the $f_x$ are uniformly distributed on $[0,1]$. The
  fitness values induce a natural orientation of the Hamming graph
  where links between neighbouring sequences are oriented in the
  direction of increasing fitness. The resulting acyclic directed
  graph is called the \textit{fitness graph} and provides a
convenient visualisation of the accessible paths \cite{JK-Crona2013,JK-deVisser2009}, see
Fig.~\ref{JK-Fig1} for illustration.

\begin{figure}[t]
\begin{center}
\includegraphics[width=0.45\textwidth]{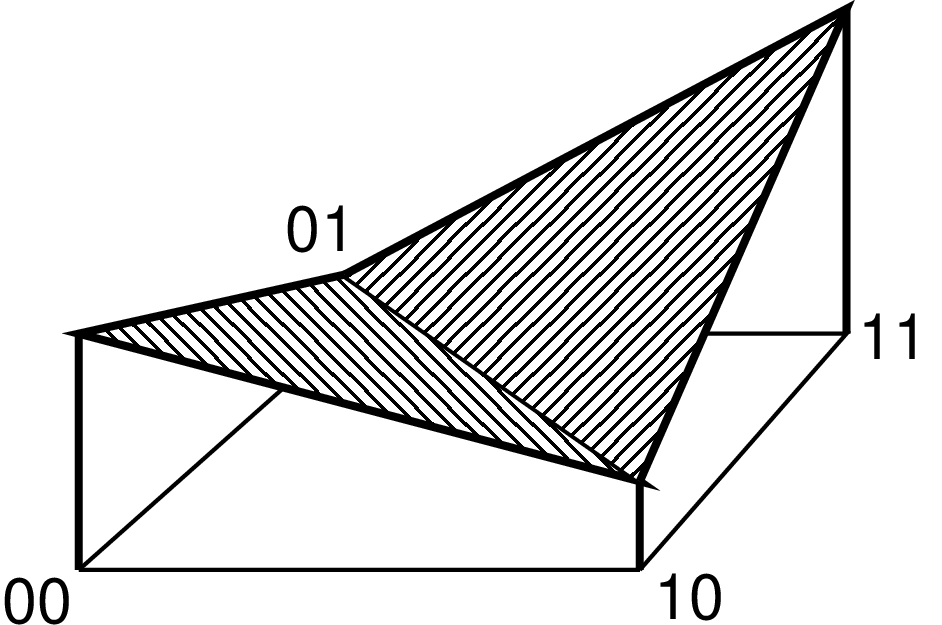} \hspace*{0.5cm} \includegraphics[width=0.4\textwidth]{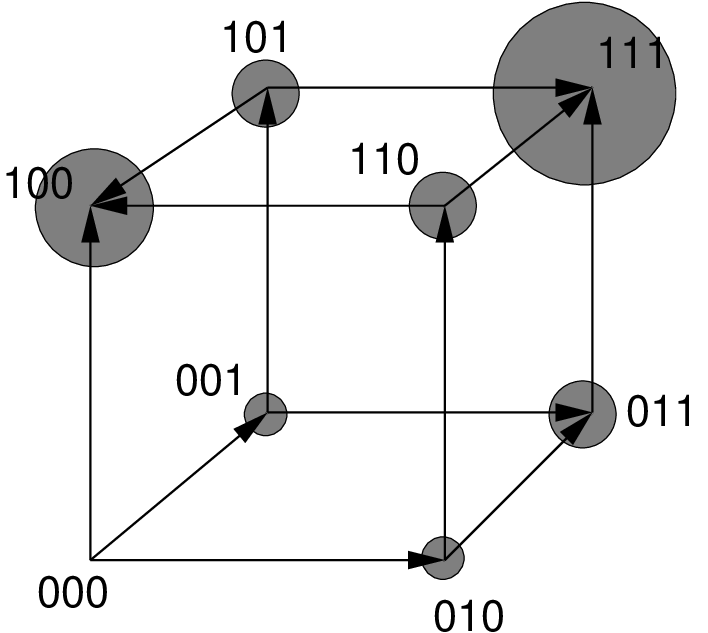}
\end{center}
\caption{\label{JK-Fig1} Fitness landscapes on the binary hypercube
  of dimension $L=2$ (left) and $L=3$ (right) loci. 
In the left panel the fitness values are plotted on the vertical axes,
whereas in the right panel they are represented by circles of different
size, and arrows on the edges point in the direction of higher fitness. The fitness
  landscape on the left is double-peaked, and as a consequence there
  are no accessible paths from $x = 00$ to $y = 11$. In the right
  panel 4 out of $6 = 3!$ possible directed paths from 000 to 111 are accessible.}
\end{figure}

The quantity of main interest in this chapter is the number of
accessible evolutionary paths between two genotypes $x,y$ with $f_y >
f_x$. Throughout we will
denote this integer-valued, non-negative random variable by 
$X_{x,y}$.
While it would be desirable to characterise
the full distribution of $X_{x,y}$, we will mostly be restricted to statements about
the expectation $\EE(X_{x,y})$ and the probability of existence of at least
one path $\PP(X_{x,y} \geq 1)$, where both quantities are
  understood to be conditioned on some positive value of the fitness
  difference $f_y - f_x$. Colloquially, we will sometimes say that a
genotype space is accessible (inaccessible) if $\PP(X_{x,y} \geq 1)
\to 1$ ($\PP(X_{x,y} \geq 1) \to 0)$ when $d(x,y) \to \infty$ \cite{JK-Franke2011}. 

\subsection{Directed paths}
\label{JK-Sec:Directed}

A path of length $\ell$ between two genotypes $x,y$ is a sequence
\begin{equation}
x \to x_1 \to x_2 \to \cdots \to x_{\ell-1} \to y
\end{equation}
of genotypes such that  
$d(x,x_1) = d(x_i,x_{i+1}) = d(x_{\ell-1},y) = 1$ for $i=1,\cdots,\ell-1$,
and it is accessible if 
\begin{equation}
\label{JK-accessible} 
f_x < f_{x_1} < f_{x_2} < \cdots < f_{x_{\ell-1}} < f_y. 
\end{equation}
Along a \textit{directed} path the distance to the 
target genotype $y$ decreases by one in each step, which implies that
$\ell = d(x,y)$ and $d(x,x_i) = i$. The properties of directed paths
do not depend on the number of alleles $a$, since the only mutations that occur
along such a path convert an allele of the initial genotype $x$ into the
corresponding allele of the final genotype $y$ \cite{JK-Zagorski2016}.
There are $\ell$ such mutations that can occur in arbitrary order, hence
the total number of directed paths is $\ell!$. 

We begin by computing the expected number of accessible directed paths
conditioned on the fitness difference $\beta = f_y - f_x > 0$. The condition
(\ref{JK-accessible}) implies that all intermediate fitness values must lie
between $f_x$ and $f_y$, which happens with probability $\beta^{\ell-1}$,
and additionally they have to be ordered, which happens with probability 
$\frac{1}{(\ell-1)!}$. Thus the probability for the path
  to be accessible is
  \begin{equation}
    \label{JK-pacc}
    P_{\beta,l} = \frac{\beta^{\ell-1}}{(\ell-1)!}.
      \end{equation}
This expression suggests an alternative interpretation
  which emphasises the link to percolation. Suppose we fix
  the initial and final fitness values at $f_x = 0$ and $f_y =1$, and
  remove all other genotypes (endowed with i.i.d. fitness values) independently with probability $1-\beta$.
Then the joint probability for a path of length $\ell$ from $x$ to $y$ to
exist \textit{and} to be accessible is precisely given by (\ref{JK-pacc}), and
$\beta$ is seen to play the role of the occupation probability in a
standard percolation problem.
      
Multiplying (\ref{JK-pacc}) with the total number
of directed paths we arrive at \cite{JK-Hegarty2014}
\begin{equation}
  \label{JK-directed-expectation}
\EE_{\ell,\beta}(X_{x,y}) = \ell! \frac{\beta^{\ell-1}}{(\ell-1)!} = 
\ell \beta^{\ell-1}.
\end{equation}
For any fixed $\beta < 1$, the expectation tends to zero
for\footnote{Here and in the following the limit $\ell \to
  \infty$ is understood to be taken along with the limit $L \to
  \infty$. In the case of directed paths we can set $\ell =
  L$ without loss of generality.} $\ell
\to \infty$, and by Markov's inequality we conclude that
$\lim_{\ell \to \infty}\PP_{\ell,\beta}(X_{x,y} \geq 1) = 0$ as
well. If $\beta$ is allowed to vary with $\ell$ as $\beta_\ell = 1 -
\epsilon_\ell$ such that $\epsilon_\ell \to 0$, the expectation
(\ref{JK-directed-expectation}) vanishes in the limit if $\epsilon_\ell > \frac{\ln \ell}{\ell}$
and diverges if $\epsilon_\ell < \frac{\ln \ell}{\ell}$. In the latter case
no statement about the existence of accessible paths can be inferred
from Markov's inequality. Hegarty and Martinsson
  \cite{JK-Hegarty2014} used the second moment inequality
  \cite{JK-Alon2000}
\begin{equation}
\label{JK-2ndmoment}
\PP(X_{x,y} \geq 1) \geq \frac{\EE(X_{x,y})^2}{\EE(X_{x,y}^2)}
\end{equation}
to show that the upper bound provided by the first
moment is essentially tight, in the sense that 
\begin{equation}
\label{JK-Hegarty-Martinsson}
\lim_{\ell \to \infty} \PP_{\ell,\beta_\ell}(X_{x,y} \geq 1) = \left\{ \begin{array}{l@{\quad:\quad}l}
 0 & \epsilon_\ell = 1 - \beta_\ell > \frac{\ln
    \ell}{\ell} + \delta_\ell \\ 
1 & \epsilon_\ell = 1 - \beta_\ell < \frac{\ln \ell}{\ell} - \delta_\ell,
\end{array} \right.
\end{equation}
where $\delta_\ell>0$ with $\lim_{\ell \to \infty} \ell
\delta_\ell = \infty$. For the estimate of the second
  moment in (\ref{JK-2ndmoment}) one needs to consider pairs of paths
  and their intersections.

Thus for directed paths a transition from low to high accessibility
occurs near $\beta = 1$, and this result can be read off from the behaviour of 
the expectation $\EE(X)$.
The full distribution of the number of accessible directed paths was
obtained by Berestycki, Brunet and Shi
\cite{JK-Berestycki2016}. Working in a scaling limit where
$\epsilon_\ell = \frac{C}{\ell}$ for $\ell \to \infty$ with $C > 0$, they show that
$X_{x,y}/\ell$ converges in law to $\exp(-C)$ times the product
of two standard independent exponential random variables. 

The result (\ref{JK-Hegarty-Martinsson}) can also be applied to the
setting originally considered in \cite{JK-Franke2011}, where the paths
were constrained to end at the global fitness maximum (which
corresponds to setting $f_y = 1$) but the initial fitness was not
specified. This amounts to integrating
(\ref{JK-directed-expectation}) with respect to $\beta = 1
  - f_x$. Remarkably,
the result
\begin{equation}
  \label{JK-unconstrained}
  \int_0^1 d\beta \, \EE(X_{\ell,\beta}) = 1
\end{equation}
is independent of $\ell$ and thus illustrates the fact that the
directed hypercube is ``marginally'' accessible. Whereas
(\ref{JK-unconstrained})  could naively be interpreted to
imply that accessible paths are likely to exist, the simulations reported in
\cite{JK-Franke2011} showed that most realisations of landscapes had $X_{x,y} =
0$, and the unit mean value was achieved through
rare instances with $X_{x,y} \gg 1$. On the basis of
(\ref{JK-Hegarty-Martinsson}) these instances are understood to
be those where the initial fitness happens to be below $\frac{\ln
  \ell}{\ell}$. 

\subsection{Paths with back steps}

From a biological perspective there is no good reason to exclude the
possibility of mutational reversions, where a mutation that occurs at
some point along the path is later reverted. Specialising to the
biallelic case with $a=2$, a path between two genotypes at distance
$D$ that includes $k$ reversions has total length $\ell = D + 2k$,
since each reversion has to be compensated by an additional forward
step. When mutations are rare, evolution is expected to proceed
preferentially along the shortest paths, and longer paths
are obviously also less likely to be accessible.
This disadvantage may however be offset by the enormous increase in
the number of possible paths, which merely have to be
self-avoiding. A re-analysis of the 5-dimensional
TEM-1 $\beta$-lactamase resistance landscape of Weinreich et al. \cite{JK-Weinreich2006} that
included mutational reversions found a moderate increase in the number
of accessible paths, from 18 to 27 \cite{JK-DePristo2007}. At the same
time the number of possible paths increases from 120 to
18,651,552,840. For hypercubes of dimension $L \geq 6$ the total number
$P_L$ of self-avoiding paths connecting two antipodal
  corners is not explicitly known, but it can be
shown to grow double-exponentially as
\cite{JK-Berestycki2017}
\begin{equation}
  \lim_{L \to \infty} \frac{\ln \ln P_L}{L} = \ln 2.
\end{equation}
This coincides with the behaviour of the naive estimate $P_L
\sim L^{2^L}$ obtained by noting that a self-avoiding path takes at
most $2^L$ steps, and that each step can proceed in $L$ different
directions. 

In the following we consider the binary hypercube of dimension $L$ and
ask for the number of general (undirected) accessible paths connecting two genotypes $x,y$ at
distance $D = d(x,y)$ that differ in fitness by $\beta = f_y - f_x > 0$.   
An expression for the expected number of paths can be
formally written down along the lines of Eq.~(\ref{JK-directed-expectation}) as
\begin{equation}
 \EE_{L,D,\beta}(X_{x,y}) = \sum_{k \geq 0} a_{L,D,k}
  \frac{\beta^{D + 2k -1}}{(D + 2k -1)!},
\end{equation}
where $a_{L,D,k}$ is the number of self-avoiding paths with $k$
reversions that connect two genotypes at distance $D$ on a hypercube of dimension $L$. Through a
careful analysis of the asymptotics of the $a_{L,D,k}$, Berestycki et
al. showed that in the joint limit $L,D \to \infty$ at fixed $\alpha =
D/L$ the exponential growth rate of the expected number
of accessible paths is given by \cite{JK-Berestycki2017}
\begin{equation}
  \label{JK-expectation-general}
  \lim_{L \to \infty} [\EE_{L,\alpha L,\beta}(X_{x,y})]^{1/L} =
  \sinh(\beta)^\alpha \cosh(\beta)^{1-\alpha}.
  \end{equation}
Similar to the directed path case discussed previously, when the right hand
side of (\ref{JK-expectation-general}) is less than 1, Markov's
inequality implies $\lim_{L \to \infty} \PP_{L,\alpha L,\beta}(X_{x,y} \geq 1) = 0$. The
condition $ \sinh(\beta^\ast)^\alpha \cosh(\beta^\ast)^{1-\alpha} = 1$ defines a
function $\beta^\ast(\alpha)$ that takes on its maximal value
$\beta^\ast(1) = \ln(1+\sqrt{2}) \approx 0.88137...$ at $\alpha = 1$
and tends to 0 for $\alpha \to 0$. Berestycki et al. conjectured that, similar to the directed case, the expectation
(\ref{JK-expectation-general}) ``tells the truth'', in the sense that
$\lim_{L \to \infty} \PP_{L,\alpha L,\beta}(X_{x,y} \geq 1) = 1$ when
$\beta > \beta^\ast(\alpha)$. This conjecture was proven independently
by Martinsson \cite{JK-Martinsson2015} and Li
\cite{JK-Li2018}. Martinsson's proof\footnote{For technical reasons
  Martinsson's proof is limited to the range $\alpha \geq 0.002$.} makes use of an ingenious
mapping to first passage percolation on the hypercube, which allows
him to refer to earlier results for the latter problem
\cite{JK-Martinsson2016}.

Thus we see that the extension to undirected paths fundamentally
changes the nature of the problem, in that the transition to high
accessibility now occurs at a nontrivial threshold fitness $\beta^\ast
< 1$. Moreover, the fact that $\beta^\ast$ decreases with decreasing
$\alpha$ shows that, in contrast to the directed path case, the
genotypes that do not lie ``between'' the initial and final point of
the path\footnote{In biological terms, these are
  genotypes that cannot be generated from the initial and final genotype
  by crossover.} cannot be ignored. Evolutionary accessibility
increases when $D$ decreases relative to $L$ because of the
contribution from paths that accumulate and later revert mutations
that are part of neither the initial nor the final genotype.   

\subsection{Multiple alleles}
\label{JK-Sec:Multiple}

In sequence spaces with more than two alleles ($a > 2$) mutational paths can
include ``sideways'' steps where the distance to the initial and final
point neither decreases nor increases, because a site mutates to an
allele that is contained in neither the initial nor the final
genotype. Zagorski et al. carried out simulations of mutational paths in
multiallelic sequence spaces and found a significant increase
of accessibility with increasing $a$ that is caused mainly by sideways steps \cite{JK-Zagorski2016}. 

In the following we summarise the main results of a recent analytic
study of this problem \cite{JK-Schmiegelt2019}. For this purpose we formalise the mutational structure on the allele set $\cA$ through
the adjacency matrix $\mathbf{A}$ of the mutation graph, with elements $A_{kk} = 0$ and $A_{kl} = 1$ iff mutations can occur from
allele $k$ to allele $l$, where $k,l \in \cA$. Then the exponential growth rate of the expected number of accessible paths between two genotypes
$x,y$ is given by 
\begin{equation}
\label{JK-expectation-multiallelic}
\lim_{L \to \infty} [\EE_{L,\mathbf{A},\beta}(X_{x,y})]^{1/L} = \prod_{k,l = 1}^{a} [(e^{\beta \mathbf{A}})_{kl}]^{p_{kl}},
\end{equation}
where $p_{kl}$ denotes the fraction of sites at which
$\tau_i(x) = k$ and $\tau_i(y) = l$ in the joint limit $d(x,y) \to \infty$ and $L \to \infty$. The information about the 
Hamming distance between $x$ and $y$ is contained in the $p_{kl}$ through the relation
\begin{equation}
\lim_{L \to \infty} \frac{d(x,y)}{L} = 1 - \sum_{k=1}^{a} p_{kk},
\end{equation} 
but in general the number of paths depends on the entire allelic composition of the initial and final genotypes. 
An important special case is the complete mutation graph, where $A_{kl} = 1 - \delta_{kl}$. In the biallelic case considered previously
this implies $\mathbf{A}^2 = \mathbf{1}$, and therefore $e^{\beta \mathbf{A}} = \sinh(\beta) \mathbf{A} + \cosh(\beta) \mathbf{1}$. Observing further that $p_{01} + p_{10} = \alpha$ and $p_{11} + p_{00} = 1 - \alpha$, the expression 
(\ref{JK-expectation-multiallelic})
is seen 
to reduce to (\ref{JK-expectation-general}).

As before, the expression (\ref{JK-expectation-multiallelic}) can be invoked together with Markov's inequality to derive a lower
bound $\beta^\ast$ on the critical fitness difference below which $\lim_{L \to \infty} \PP_{L,\mathbf{A},\beta}(X_{x,y} \geq 1) = 0$. For the 
complete mutation graph over $a$ alleles and initial and final genotypes at maximal distance $d=L$, the equation for 
$\beta^\ast(a)$ reads
\begin{equation}
\label{JK-betaast}
\frac{1}{a} \left( e^{(a-1) \beta^\ast} - e^{-\beta^\ast} \right) = 1,
\end{equation}
which can be solved explicitly for $a \leq 4$. In particular, for the case of the 4-letter nucleotide alphabets of RNA and DNA
the threshold fitness is 
\begin{equation}
\beta^\ast(4) = \ln \left( \frac{1}{\sqrt{2}} + \sqrt{(\sqrt{2} - \frac{1}{2})} \right) \approx 0.5088...
\end{equation} 
For large $a$ the solution of (\ref{JK-betaast}) can be approximated as 
\begin{equation}
\label{JK-betaast-expansion}
\beta^\ast(a) = \frac{\ln(a)}{a} + \frac{1+\ln(a)}{a^2} + {\mathcal{O}}\left(\frac{\ln(a)}{a^3}\right).
\end{equation}
For the 20-letter amino acid alphabet this yields the estimate $\beta^\ast \approx 0.1598...$. Because of the restrictions of the 
genetic code, the amino acid mutation graph is not complete, but a calculation based on the actual mutation 
graph shows that this only leads to minor deviations from this estimate.  

The comparison to numerical simulations \cite{JK-Zagorski2016} and 
related results for first passage percolation \cite{JK-Martinsson2018}  
indicate that the bound $\beta^\ast$ given by setting the right hand side of (\ref{JK-expectation-multiallelic}) to unity is tight at least
for the complete graph, and presumably also for a rather general class
of allelic mutation graphs. An interesting application for which the lower bound
provided by (\ref{JK-expectation-multiallelic}) suffices is the
linear graph\footnote{A possible
  biological interpretation of the linear mutation graph is that alleles represent
  copy-number variants of genes \cite{JK-Altenberg2015}. In this case the
assignment of random fitness values is however not very plausible.}, where allele $k$ is allowed to mutate only to 
the neighbouring alleles $k \pm 1$ for $1 \leq k \leq a-2$, and the
boundary alleles mutate as $0 \to 1$ and $a-1 \to a-2$. For paths connecting the boundary
genotypes $x,y$ with $\tau_i(x) = 0$ and $\tau_i(y) = a-1$ or vice versa,
one finds that $\beta^\ast > 1$ for $a > 2$, which implies that accessible 
paths do not exist for any value of $\beta \in [0,1]$.

\section{Accessibility percolation on trees}
\label{JK-Sec:Trees}

\setcounter{footnote}{0}

Computing higher moments of $X_{x,y}$ on the directed hypercube is difficult
because different paths can merge and diverge multiple times
\cite{JK-Hegarty2014}. The observation that sequence spaces are nevertheless essentially
tree-like for large $L$ has motivated a number of studies of
accessibility percolation on trees, where this problem does not arise.  


We begin with a regular rooted $n$-tree of height $h$
\cite{JK-Nowak2013}. The tree has $n^h$ leaves and equally many paths
of length $h+1$ from the root to one of the leaves. The nodes and leaves
are labelled by continuous, independent and identically distributed (i.i.d.)
random numbers, and therefore 
the probability for a given path to be accessible is $1/(h+1)!$. Since
the exponential growth of the number of paths cannot compensate the
factorial decrease in probability, the usual first moment bound shows
that accessible paths do not exist for $h \to \infty$ for any fixed
value of $n$. One is thus lead to consider trees where the branching number grows with
increasing height according to a function $n(h)$. The expected number
of accessible paths is then\footnote{Throughout this
  section paths are assumed to go from the root to a leaf of the
  tree. The indices $x,y$ of $X$ indicating the start and end point of the
  paths are therefore omitted.} 
\begin{equation}
  \EE_{h,n(h)}(X) = \frac{n(h)^h}{(h+1)!} \sim \frac{[e
    n(h)/h]^h}{\sqrt{2\pi h}}
\end{equation}
for large $h$,
which suggests that the transition to high accessibility occurs for linear
functions $n(h) = \lambda h$ with $\lambda > \lambda^\ast \geq
1/e$. In \cite{JK-Nowak2013} the upper bound $\lambda^\ast \leq 1$ was obtained
using the second moment inequality (\ref{JK-2ndmoment}), and a subsequent
refined analysis showed that the lower bound is tight and $\lambda^\ast =
1/e$ exactly \cite{JK-Roberts2013}.
The linear growth $n \sim h$
corresponds to the geometry of the directed hypercube\footnote{See
  \cite{JK-Berestycki2016} for a precise way of approximating the
  directed hypercube by a tree.}, which can be viewed as
a directed graph whose vertex degree and diameter are both equal to
$L$. The fact that trees with linear growth are marginally accessible
is thus consistent with the results for the directed hypercube described in
Sec.~\ref{JK-Sec:Directed}.

A similar conclusion can be drawn from a subtly different
analysis carried out by Coletti et al. \cite{JK-Coletti2018}, who
consider infinite trees with the branching number at level $l$ given
by an increasing function $n(l)$ (the root is located at
$l=0$). For a linear growth function $n(l) = l+1$ the number of leaves (and hence the number of
distinct paths from the root) at height $h$ is $h!$, the same as the
number of directed paths on a hypercube of dimension $h$. Without
constraints on the fitness of the root the probability for a path to
be accessible is again $1/(h+1)!$. The expected number
of paths is  
\begin{equation}
\EE_{h,n(l) = l+1}(X) = \frac{h!}{(h+1)!} = \frac{1}{h+1} 
\end{equation}
and there is no accessibility for $h \to \infty$. The main result
established in \cite{JK-Coletti2018} is that this case is marginal in
the sense that the probability for existence of accessible paths is
positive for growth functions $n(l) = \lceil (l+1)^\gamma \rceil$ with
$\gamma > 1$. 

Instead of letting the branching number of the tree grow with its
height, accessibility can also be increased by introducing a bias on
the random fitness variables \cite{JK-Nowak2013}. Specifically, we
take the fitness of a node $x$ at
distance $l$ from the root to be of the form
\begin{equation}
  \label{JK-tree}
  f_x = \xi_x + c l
\end{equation}
where $c > 0$ and the $\xi_x$ are continuous i.i.d. random variables. The linear trend in (\ref{JK-tree}) increases the likelihood of the variables to be 
in increasing order in a way that depends on the distribution of the $\xi_x$. 
For the case of the Gumbel distribution $\PP(\xi_x < z) =
\exp[-e^{-z}]$, the ordering probability $\PP(f_{x_1} <
  f_{x_2} < \cdots < f_{x_h})$, 
which is also the probability for a path of length $h$ to be accessible, can be shown to be given by \cite{JK-Franke2010}
\begin{equation}
  \label{JK-ordering}
  \PP(f_{x_1} < f_{x_2} < \cdots < f_{x_h}) =
  \frac{(1-e^{-c})^h}{\prod_{l=1}^h (1-e^{-cl})} =
  \frac{1}{[h]_{e^{-c}}!}.
\end{equation}
Here
\begin{equation}
  [k]_q! \equiv \prod_{j=1}^k [j]_q, \;\;\; [j]_q \equiv
  \frac{1-q^j}{1-q}, \;\;\; q \in [0,1], 
\end{equation}
defines the $q$-factorial of a $q$-number\footnote{For a related
  application of $q$-factorials see \cite{JK-Park2016}.} $[k]_q$
\cite{JK-Koekoek2010}. Note that for $q \to 1$ the standard factorial
is retrieved. The result (\ref{JK-ordering}) was first derived in the context of
record statistics, where it describes the probability for all entries
in a sequence of random variables with a linear trend to be
records\footnote{A similar relation between record statistics and
  accessibility was described in \cite{JK-Coletti2018}.}
\cite{JK-Franke2010}. 

Since the product in the denominator of (\ref{JK-ordering}) converges
for $h \to \infty$, for a tree of fixed branching number $n$, the
expected number of accessible paths grows or shrinks
exponentially as $[n(1-e^{-c})]^h$ for large $h$. This suggests that
the accessibility transition occurs when $n(1-e^{-c}) = 1$ or $c =
c^\ast$ with
\begin{equation}
  c^\ast = \ln\left(\frac{n}{n-1}\right). 
\end{equation}
An analysis using the second moment inequality (\ref{JK-2ndmoment}) confirms this
expectation and moreover provides the lower bound
\begin{equation}
 \lim_{h \to \infty} \PP_{h,n,c}(X \geq 1) \geq \sqrt{\frac{2\pi}{c}}
   \exp \left( \frac{c}{24}-\frac{\pi}{6c} \right)[e^{-c^\ast}-e^{-c}]
\end{equation}
for $c > c^\ast$ \cite{JK-Nowak2013}.

\section{Correlated fitness landscapes}
\label{JK-Sec:Correlations}

Although the assumption of i.i.d. random fitness values is
mathematically convenient, it cannot be expected to be biologically
realistic. Indeed, analyses of experimental data
generally show that real fitness landscapes are less rugged than
predicted by the i.i.d. model
\cite{JK-Bank2016,JK-deVisser2014,JK-Franke2011,JK-Szendro2013}.
For this reason several classes of probabilistic
fitness landscapes with tunable fitness
correlations have been proposed. In contrast to the i.i.d. case, in these models the rank ordering of genotypes
(and hence the accessibility of mutational paths) generally depends on the base distribution of the random variables from which 
the landscape is constructed. In the following we focus on properties for which this dependence does not matter, and consider
biallelic sequence spaces ($a = 2$) throughout. 

In the rough Mount Fuji (RMF) model, i.i.d. random fitness
values are combined additively with a linear\footnote{For a
  generalisation to nonlinear functions see \cite{JK-Park2015,JK-Park2016a}.} function of
the distance to a reference genotype $x_0$, such that
\begin{equation}
\label{JK-RMF}
  f_x = \xi_x + c d(x,x_0)
\end{equation}
with $c > 0$ and continuous i.i.d. random variables $\xi_x$
\cite{JK-Neidhart2014}. This is obviously very similar to the fitness
assignment (\ref{JK-tree}) in the tree
model with bias discussed in Sec.~\ref{JK-Sec:Trees}. The RMF model on
the directed hypercube was analysed by Hegarty and Martinsson
\cite{JK-Hegarty2014}, who showed that accessible paths of length $L$
starting at $x_0$ exist with probability converging to 1 for any $c > 0$ and $L
\to \infty$. The effect of the bias in the undirected case
is less clear cut, since mutational reversions are discouraged when $c >
0$. In particular, for $c \to \infty$ all directed paths become
accessible and no back steps can occur. Numerical simulations for small
systems suggest that the competition between directed and undirected
paths leads to a maximum in the number of accessible paths at an
intermediate value of $c$ \cite{JK-Josupeit2015}. 

The NK-models\footnote{The acronym refers to the number of
  loci $N$ (here denoted by $L$) and the number of
  interaction partners $K$ of a locus (here denoted by $k-1$).} originally introduced by Kauffman and Weinberger \cite{JK-Kauffman1989,JK-Kauffman1993} constitute a popular and versatile class of correlated 
fitness landscapes that continues to attract the attention
of diverse research communities (see \cite{JK-Hwang2018} for a recent review). The basic structural
element of the model are the \textit{interaction sets} $B_i \subset \cL$ of genetic loci, which 
are subsets of the locus set $\cL = \{1,2,\cdots,L\}$. 
In the most commonly used setting there are $L$ 
interaction sets of equal size $\vert B_i \vert = k$, $1 \leq k \leq L$, and moreover it is assumed that locus $i$ belongs
to its own interaction set, $i \in B_i$. Loci within one interaction set affect each others fitness effects in a random manner, 
whereas the fitness effects from different interaction sets combine additively. This is implemented by writing the fitness of a genotype
$x$ as a sum over contributions from the interaction sets, 
\begin{equation}
\label{JK-NK}
f_x = \sum_{i=1}^L \phi_k^{(i)}(\downarrow_{B_i}x).
\end{equation}
Here the $\phi_k^{(i)}$ are functions on the $k$-dimensional hypercube which assign a continuous i.i.d. random number to each
of the $2^k$ sequences in  $\HH_2^k$, and $\downarrow_{B_i}:\HH_2^L \to \HH_2^k$ projects the genotype sequence onto the interaction
set according to 
\begin{equation}
\tau_i(\downarrow_{\cS}x) = \tau_i(x) \;\; \forall i \in \cS \subset \cL.
\end{equation}    
The functions $\phi_k^{(i)}$ for different $i$ are independent. The choice of the interaction sets defines the 
\textit{interaction structure} of the model \cite{JK-Hwang2018,JK-Nowak2015}, see Fig.~\ref{JK-Fig2} for illustration. 

\begin{figure}[t]
\begin{center}
\includegraphics[width=0.3\textwidth]{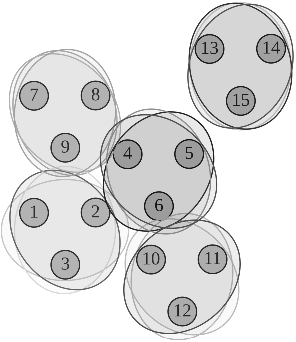} \hspace*{0.2cm} \includegraphics[width=0.3\textwidth]{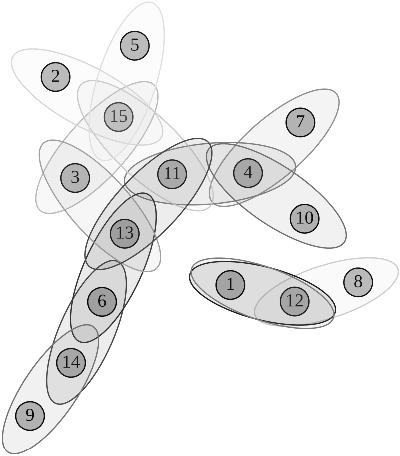} \hspace*{0.2cm} 
\includegraphics[width=0.3\textwidth]{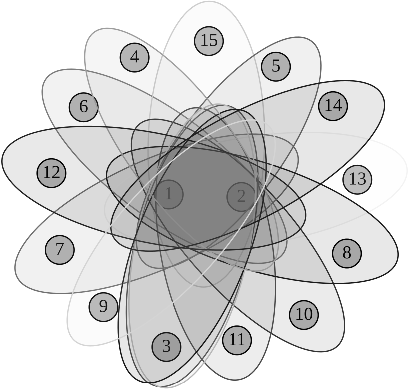}
\end{center}
\caption{\label{JK-Fig2} Three examples of NK interaction structures for $L=15$ loci. Interaction sets are shown as ellipses. 
Left panel: Block structure \cite{JK-Orr2006,JK-Perelson1995} with $k=3$. The interaction sets are disjoint and form five triples of
identical sets. Middle panel: Random structure with $k=2$. Loci are assigned randomly to interaction sets subject to the constraints
described in the text. Right panel: Star structure with $k=3$. Loci 1 and 2 are \textit{center loci} that are contained in all interaction sets. 
The interaction sets associated with the remaining 13 \textit{ray loci} contain only the locus itself and the center loci (modified from
\cite{JK-Hwang2018}).}
\end{figure}

The correlations in the fitness landscape are tuned through the size
$k$ of the interaction sets \cite{JK-Campos2002}. For $k = L$ every interaction set 
contains all loci and the model reduces to the uncorrelated landscape. On the other hand, when $k=1$ the fitness effects of different
loci combine additively according to 
\begin{equation}
f_x = \sum_{i=1}^L \phi_1^{(i)}(0) + \sum_{i=1}^L (\phi_1^{(i)}(1)-\phi_1^{(i)}(0)) \tau_i(x),
\end{equation}
where the $\phi_1^{(i)}(0)$ and $\phi_1^{(i)}(1)$ are continuous
i.i.d. random variables.
The additive landscape has a unique maximum and all directed paths to the maximum are accessible \cite{JK-Weinreich2005}. 

For general $k$, the analysis of directed accessible paths is relatively straightforward for the block structure shown in the 
left panel of Fig.~\ref{JK-Fig2} \cite{JK-Schmiegelt2014}. Note first that in this case the $L$ interaction sets fall into
$b = L/k$ groups\footnote{We assume here that $L$ is divisible by $k$.} 
within which the sets are identical, and therefore only $b$ interaction sets have to be distinguished.
Following the setting originally introduced in \cite{JK-Franke2011}, we 
consider directed paths of length $L$ that end at the global maximum $x_\mathrm{max}$ 
of the fitness landscape and start at its antipodal point $\overline{x}_\mathrm{max}$
defined by $\tau_i(\overline{x}_\mathrm{max}) = 1 - \tau_i(x_\mathrm{max})$. 
Along such a path each locus has to be mutated once. Since the interaction sets are disjoint, each mutation event occurs in one of 
the sets and changes only the fitness contribution corresponding to this set (compare to (\ref{JK-NK})). In this way the path can be 
decomposed into $b$ sub-paths of length $k$ on $\HH_2^k$, each of which remains within one interaction set. 
The global path is accessible iff all the sub-paths are. Denoting the number of accessible paths in the interaction set $B_i$ by 
$X_{k}^{(i)}$, the number of global paths is thus 
\begin{equation}
\label{JK-blockpaths}
X_{L,k}^\mathrm{block} = \frac{L!}{(k!)^b} \prod_{i=1}^b X_{k}^{(i)}.
\end{equation}
The combinatorial prefactor describes the number of ways in which a given set of sub-paths can be combined into a global path. 
An immediate consequence is that $X_{L,k}^\mathrm{block}$ is a non-negative multiple of $\frac{L!}{(k!)^b}$.

Since the fitness contributions within each interaction set are continuous i.i.d. random variables, the statistics of the $X_k^{(i)}$ 
are given by the uncorrelated model discussed in Sect.~\ref{JK-Sec:Directed}. In particular, it follows from (\ref{JK-unconstrained}) 
that $\EE(X_k^{(i)}) = 1$ independent of $k$, and therefore 
\begin{equation}
\label{JK-blockexpectation}
\EE(X_{L,k}^\mathrm{block}) = \frac{L!}{(k!)^b}.
\end{equation}     
Similarly, 
\begin{equation}
\PP(X_{L,k}^\mathrm{block} \geq 1) = [\PP(X_k \geq 1)]^b,  
\end{equation}
where $X_k$ is the number of accessible directed paths in the uncorrelated model on $\HH_2^k$ 
that end at the global maximum and start at the (unconstrained) antipodal point. The results described in Sec.~\ref{JK-Sec:Directed}
imply that\footnote{Specifically, $\PP(X_2 \geq 1) = 2/3$ and  $\PP(X_3 \geq 1) = 97/210 \approx 0.462...$ \cite{JK-Schmiegelt2014}.}  
$\PP(X_k \geq 1) < 1$ for any $k \geq 2$, and we conclude that 
\begin{equation}
\label{JK-blockaccessibility}
\lim_{L \to \infty} \PP(X_{L,k}^\mathrm{block} \geq 1) = 0  
\end{equation}
for any fixed $k \geq 2$. Taken together, the results (\ref{JK-blockexpectation}) and 
(\ref{JK-blockaccessibility}) show that the accessibility properties of the NK
model with block interactions are strikingly different from those of the uncorrelated model, in that the expected number of accessible
paths grows factorially with $L$, while at the same time the
probability that the landscape is accessible vanishes exponentially.
It is clear from the second moment inequality (\ref{JK-2ndmoment}) that these two statements are compatible only if the coefficient
of variation of $X_{L,k}^\mathrm{block}$ diverges with $L$. This is indeed the case and follows from the multiplicative structure
of (\ref{JK-blockpaths}) \cite{JK-Schmiegelt2014}. 

An early numerical investigation indicated that the accessibility of 
NK landscapes depends on the interaction structure in a significant and 
complicated way \cite{JK-Franke2012}, but subsequent work has shown that
the behaviour of the block structure described above is
quite representative at least asymptotically for large $L$. In brief, it has been proved that for a large class
of interaction structures characterised as \textit{locally bounded}, the probability 
$\PP(X_{L,k} \geq 1)$ vanishes exponentially in $L$ for $L \to \infty$ 
and any fixed $k \geq 2$ \cite{JK-Hwang2018,JK-Schmiegelt2016}. The proof 
relies on the inevitable existence of a certain local genetic interaction 
motif known as reciprocal sign epistasis \cite{JK-Poelwijk2007} which prevents
an accessible path (directed or undirected) from traversing the hypercube
and hence decomposes the genotype space into mutually inaccessible domains. 
Most commonly studied interaction structures, including the random structure 
depicted in the middle panel of Fig.~\ref{JK-Fig2} are locally bounded, but the star structure shown in the right panel of 
the figure is not.  


\section{Paths with valley crossing}
\label{JK-Sec:Downhill}

\setcounter{footnote}{0}

In this section we consider the consequences of relaxing the condition of strict monotonicity on accessible paths.
Theoretical studies of populations that cross a fitness valley have shown that different dynamic modes for this process
have to be distinguished \cite{JK-Weinreich2005a}. In small
populations\footnote{The condition on the population size $N$ 
is $N \Delta f \sim 1$, where $\Delta f = f_{x} - f_v > 0$ is the fitness difference between the initial genotype $x$ and the valley
genotype $v$.} a deleterious mutation that decreases fitness
can fix, which implies that the entire population moves to a state of lower fitness and continues its trajectory from
there. By contrast, in larger populations a small sub-population of valley genotypes is maintained by mutation-selection balance, and if 
a target genotype that is fitter than both the initial and the valley genotypes exists, 
it can arise by mutation from the valley population.  
In the latter case the majority of the population never resides in the valley, and for this reason the process is also known 
as \textit{stochastic tunnelling} \cite{JK-Iwasa2004}. Increasing
population size even further, eventually double mutations become
likely. In the following we focus on the regime where multiple
mutations can be neglected and valleys are crossed either by fixation
or stochastic tunnelling.   

\begin{figure}[t]
\begin{center}
\includegraphics[width=0.7\textwidth]{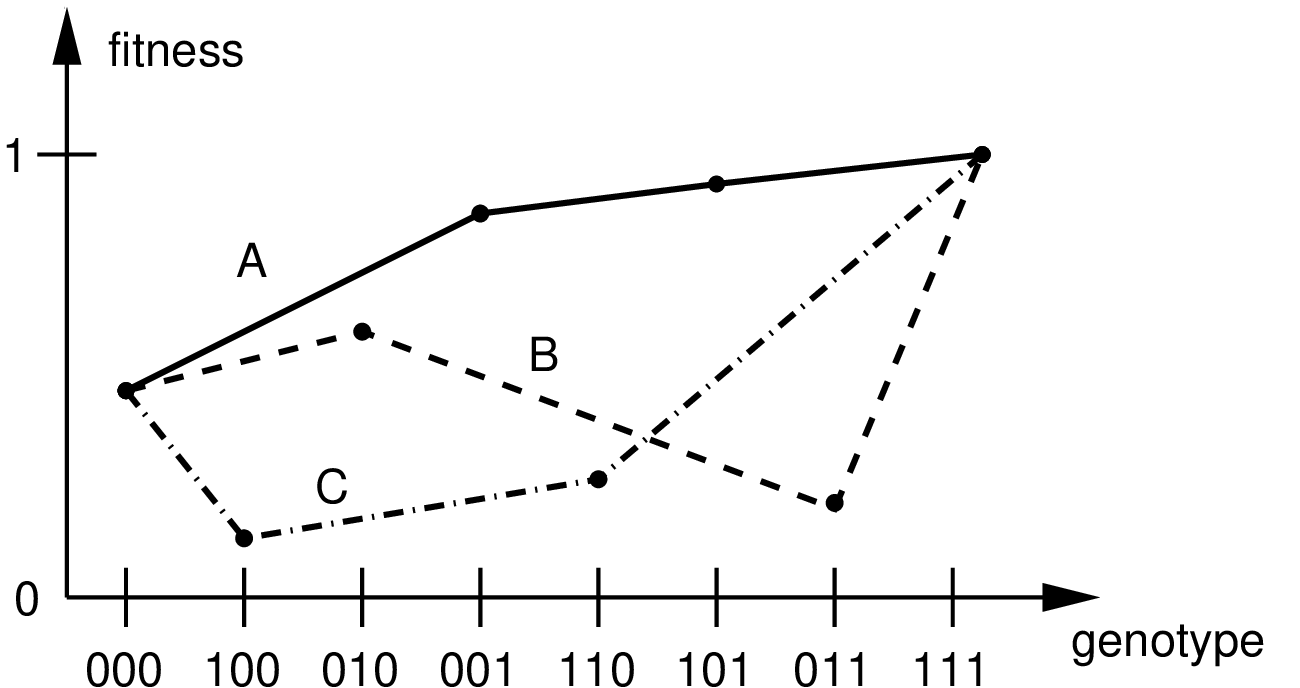}
\end{center}
\caption{\label{JK-Fig3} Three directed paths in a three-dimensional hypercube that illustrate the different modes of valley crossing. Path A (full) is monotonic, path B (dashed) contains a stochastic tunnelling event in the second step, and along path C (dash-dotted) a deleterious mutation
fixes in the first step.}
\end{figure}

These two modes of valley crossing impose different conditions on the fitness values along the evolutionary trajectory 
\cite{JK-Deecke2015}. For the \textit{fixation mode} the crossing event implies an unconditional decreases of the fitness values, after 
which the monotonic increase in fitness is resumed until the next valley crossing. By contrast, when valley crossing occurs in the 
\textit{(stochastic) tunnelling} mode, the mutational step out of the valley has to overcompensate the fitness decrease of the preceding 
step. Denoting the valley genotype by $v$, the fitness values along the sub-trajectory $x \to v \to x'$ have to satisfy the condition
$f_{x'} > f_x > f_v$, see Fig.~\ref{JK-Fig3} for illustration. In the following we examine the effect of allowing for one or 
several valley crossings on the accessibility of directed paths of length $\ell$ on the hypercube. 
Fitness values are drawn independently from the 
uniform distribution on $[0,1]$, and the final genotype $y$ is assumed to have maximal fitness, $f_y = 1$.  

We first show that a single valley crossing of the fixation type suffices to induce high accessibility. Starting at an arbitrary
initial genotype $x$ with fitness $f_x$, we make use of the crossing event in the first step to move to the genotype $x_1$ 
that has the lowest fitness among the $\ell$ available neighbours. The
probability distribution function of this fitness value is given by 
\begin{equation}
\PP(f_{x_1} < z) = 1 - (1-z)^\ell \to 1-e^{-z\ell} 
\end{equation}  
for large $\ell$ and small $z$, 
and using (\ref{JK-Hegarty-Martinsson}) we conclude that an accessible path will be found with certainty from the second step onward. 
The expected number of accessible paths starting from a random fitness
value can be shown to grow exponentially as $2^\ell-\ell$ in this 
setting \cite{JK-Deecke2015}. 

The analysis of paths with stochastic tunnelling events is more subtle\footnote{This part is based on unpublished notes by 
\'{E}ric Brunet \cite{JK-Brunet2015}.}. Following the setup of Sect.~\ref{JK-Sec:Directed}, 
we fix the initial and final fitness values at $f_x = 1-\beta$ and $f_y = 1$, respectively. We consider paths of length $\ell$ 
with a fixed number $t$ of tunnelling events, and denote by $\psi_t(\ell,\beta)$ the probability that such a path is accessible. 
We know from (\ref{JK-directed-expectation}) that $\psi_0(\ell,\beta) = \frac{\beta^{\ell-1}}{(\ell-1)!}$. For general $t$  and 
$\ell \geq 3$, the $\psi_t$ satisfy the recursion relation 
\begin{equation}
\label{JK-recursion}
\psi_t(\ell,\beta) = (1-\beta) \int_{1-\beta}^1 du \, \psi_{t-1}(\ell-2,u) + \int_{1-\beta}^1 du \, \psi_t(\ell-1,u).
\end{equation}
The first term on the right hand side describes the situation when the first step is a tunnelling event. This implies that the 
second fitness value $f_{x_1} \in [0,1-\beta]$, which happens with probability $1-\beta$, whereas the subsequent 
value $u = f_{x_2}$ has to be in $[1-\beta,1]$. From this point on one has to cover the remaining $\ell-2$ steps using $t-1$ 
tunnelling events. The second term covers the cases where fitness increases in the first step, 
such that $f_{x_1} = u \in [1-\beta,1]$ and the remaining $\ell-1$ steps can make use of all $t$ tunnelling events. The solution of 
(\ref{JK-recursion}) with the appropriate boundary conditions is given by    
\begin{eqnarray}
\label{JK-recursion-solution}
\psi_t(\ell,\beta) & = & \frac{1}{2^t t! (\ell-2t)!} \beta^{\ell-t}(2-\beta)^t \left[\frac{\ell-t}{\beta} - \frac{t}{2-\beta} \right] \\
& = & \frac{1}{2^t t! (\ell-2t)!} \frac{d}{d\beta} \left[ \beta^{\ell-t}(2-\beta)^t \right]. \nonumber
\end{eqnarray} 
Integrating (\ref{JK-recursion-solution}) with respect to $\beta$ and 
multiplying by the total number of directed paths $\ell!$ one obtains the expected number of paths with arbitrary starting fitness as 
 \begin{equation}
\EE(X_{\ell,t}) = \frac{\ell!}{2^t t! (\ell-2t)!} \,
\propto \, \ell^{2t}
\end{equation}
for large $\ell$ and fixed $t$,
which generalises (\ref{JK-unconstrained}) to $t \geq 1$. 
Although this suggests a significant increase in accessibility, the algebraic growth in 
$\ell$ is not sufficient to overcome the exponential reduction in probability caused by the factor $\beta^{\ell-t}$ in
(\ref{JK-recursion-solution}). Conditioned on the initial fitness, the expected number of paths behaves as  
\begin{equation}
\EE(X_{\ell,t,\beta}) \, \propto \,
\frac{\ell^{2t+1}}{2^t t!} \beta^{\ell-t-1}(2-\beta)^t \,
\propto \, \ell^{2t+1}\beta^\ell
\end{equation}
for large $\ell$ and fixed $t$, which converges to zero for any fixed $\beta < 1$. Setting $\beta = \beta_\ell = 1 - \epsilon_\ell$ with 
$\lim_{\ell \to \infty} \epsilon_\ell = 0$, we see that a necessary
condition for accessibility is $\epsilon_\ell < (2t+1)\frac{\ln \ell}{\ell}$, which is only a 
small improvement compared to the case without valley crossings in (\ref{JK-Hegarty-Martinsson}). 
We conclude that the effects of the two modes of valley crossing on accessibility are very different. 

Duque et al. recently studied accessibility percolation with valley crossings on $n$-trees with a height-dependent 
branching number $n(h)$, compare to Sec.~\ref{JK-Sec:Trees}. 
They define a path to be $k$-accessible if any $k$ consecutive fitness values along the path contain
at least one element of a monotonically increasing sub-sequence, and show that the critical growth of $n(h)$ required
to guarantee accessibility is given by $n(h) \, \propto
\, [h/(ek)]^{1/k}$ \cite{JK-Duque2019}.

\section{Summary and conclusions}
\label{JK-Sec:Summary}

\subsection{Accessibility and predictability}

The results reviewed in the preceding sections reveal a variety of mathematical mechanisms by which evolutionary 
accessibility can arise in high-dimensional 
genotype spaces. The directed hypercube with continuous i.i.d. fitness values  
\cite{JK-Carneiro2010,JK-Franke2011} turns out to be marginally accessible, in the sense that accessible paths exist only if
the fitness difference between the initial and final genotype is as large as possible \cite{JK-Hegarty2014}. 
Allowing for undirected paths
lowers the threshold for accessibility to a nontrivial value $\beta^\ast \in (0,1)$ \cite{JK-Berestycki2017,JK-Li2018,JK-Martinsson2015}. 
In multiallelic sequence spaces $\beta^\ast$
depends on the graph of allowed mutational transitions on the set of alleles $\cA$. If the mutation graph is complete, $\beta^\ast$
decreases with increasing number of alleles and tends to zero for $a \to \infty$ \cite{JK-Schmiegelt2019,JK-Zagorski2016}. 
When the genotype space is a tree, accessibility percolation occurs as a function of tree geometry 
\cite{JK-Coletti2018,JK-Nowak2013,JK-Roberts2013} or as a function of 
a bias imposed on the fitness values \cite{JK-Nowak2013}. Apart from the last example, in all cases 
the transition is characterised by a discontinuous change of the limiting value of $\PP(X \geq 1)$ from 0 to 1. 

On correlated fitness landscapes of NK-type, accessibility is determined by the graph of interactions among loci, and accessible 
paths do not exist for interaction structures that are locally bounded \cite{JK-Hwang2018,JK-Schmiegelt2016}.  
This general result as well as the results for the explicitly solvable block model \cite{JK-Schmiegelt2014} show that,
despite being less rugged (more correlated) than the uncorrelated model, NK-landscapes are much less accessible. 
Evolutionary accessibility is thus not strictly linked to other measures of fitness landscape ruggedness such as 
the number of local fitness maxima \cite{JK-Szendro2013}. 

The concept of accessible paths was originally introduced in the
context of evolutionary predictability. 
Following the interpretation that Weinreich et al. applied to their seminal experiment on antibiotic resistance evolution
\cite{JK-Weinreich2006}, the evolutionary trajectories connecting an initial to a final genotype are highly predictable 
if a small but nonzero number of accessible pathways exist. Referring to the results described in Sect.~\ref{JK-Sec:SequenceSpace}, 
we see that this condition is approximately satisfied for the directed hypercube with $\beta = 1$, where $\lim_{L\to\infty} \PP(X \geq 1) = 1$ and the expected number of paths is equal to $L$ and hence much smaller than the total number of paths $L!$. For the undirected
hypercube at $\beta > \beta^\ast$ the number of accessible paths increases exponentially in $L$ but is still much smaller than 
the total number of paths. In this sense the uncorrelated model can be said to conform to the scenario of high predictability 
envisioned in \cite{JK-Weinreich2006}.
However, the results for the NK-models discussed in Sect.~\ref{JK-Sec:Correlations} show that other scenarios are possible as well.
According to Eqs~(\ref{JK-blockexpectation},\ref{JK-blockaccessibility}), 
in the NK-model with block interactions accessible paths typically do not exist, but if they do
their number grows factorially in $L$, leading to low predictability. 

\subsection{Accessible paths and evolutionary dynamics}
\label{JK-Sec:Dynamics}

Formalising evolutionary accessibility through fitness-monotonic mutational paths is conceptually appealing, because
it does not require any assumptions about the evolutionary dynamics. On the other hand, this simplification also limits the applicability
of the theory to actual evolutionary processes. The paradigm of evolutionary dynamics 
underlying the notion of accessible paths, known as the strong
selection-weak mutation (SSWM) regime, applies in a window of 
population size $N$ where mutations are rare\footnote{Here $\mu$ denotes the mutation rate per individual and generation, 
and $\Delta f$ is the typical fitness difference between neighbouring genotypes.}, 
$N \mu \ll 1$, and selection is strong, 
$N \vert \Delta f \vert \gg 1$ \cite{JK-Gillespie1984,JK-Orr2002}. Under these conditions the population
is almost always monomorphic and can evolve only by fixing beneficial
mutations, which implies that it is constrained to move along
accessible paths. A rigorous derivation of the SSWM limit from microscopic
adaptive dynamics is described in the chapter by Anton Bovier \cite{JK-Bovier2019}. 
The SSWM framework allows one to assign a weight to each accessible path, which is given by the product of the relative
fixation probabilities along the path. Empirical studies indicate that these weights are often strongly concentrated
on a small number of paths \cite{JK-Lozovsky2009,JK-Weinreich2006}.
As a consequence the evolutionary predictability may be even higher than expected
based only on the number of accessible paths.   

However, evolving populations are myopic and lack the foresight required to determine which of the many available paths will eventually
lead to high fitness. Studies of SSWM dynamics on the $L$-dimensional
hypercube with uncorrelated fitness values have shown that adaptive walks
governed by local dynamics typically terminate
at a local fitness peak after $\sim \ln L$ steps and are thus much shorter than the accessible paths considered in this chapter
\cite{JK-Flyvbjerg1992,JK-Macken1991,JK-Neidhart2011,JK-Orr2002}. 
Greedy adaptive walks that always fix the most beneficial mutation terminate already after $e-1 \approx 1.718$ steps
\cite{JK-Orr2003}. Walks following a (biologically unrealistic) `reluctant' dynamics by always choosing the smallest available positive fitness difference take $\cO(L)$ steps but do not reach fitness levels comparable to the globally maximal fitness \cite{JK-Nowak2015}. 
The behaviour of adaptive walks on correlated fitness landscapes is more complex, but studies of the RMF model have shown that 
the available, long accessible paths are dynamically relevant only if the bias parameter $c$ in (\ref{JK-RMF}) is sufficiently
large and/or the distribution of the random component $\xi_x$ is sufficiently heavy tailed 
\cite{JK-Park2015,JK-Park2016,JK-Park2016a}. Typical walk lengths in the
NK-model at fixed $k$ are of the order of (but smaller than) $L$
\cite{JK-Nowak2015}. 

When the population size increases beyond the weak mutation regime additional complications arise.
On the one hand, the simultaneous presence of multiple mutation clones in the population implies an advantage for mutations 
of large beneficial effect, such that the dynamics becomes
increasingly greedy and hence deterministic \cite{JK-Jain2011,JK-Park2016}. 
On the other hand, a higher mutation
rate facilitates the crossing of fitness valleys, which strongly
increases the number of available paths.
A detailed numerical study has shown that the distinguished role of 
fitness-monotonic paths is largely lost in this regime, and moreover the
interplay of the two effects mentioned above leads to a non-monotonic dependence of predictability on population size 
\cite{JK-Szendro2013a}.

Taken together, the considerations sketched in this section make it clear that the investigation
of accessible paths only constitutes a first step towards a broader understanding of evolutionary predictability. 

\vspace*{0.3cm}

\noindent
\textbf{Acknowledgement.} The work described here is the joint effort of a large number of collaborators. I am particularly
grateful to \'{E}ric Brunet, Lucas Deecke, Jasper Franke, Mario Josupeit, Alexander Kl\"ozer, Stefan Nowak and
 Benjamin Schmiegelt for their contributions to this project, and to
 Benjamin Schmiegelt for a critical reading of the manuscript.

\end{document}